\documentclass[12pt]{article}
\pdfoutput=1
\usepackage{geometry,enumerate,amsmath,amssymb}
\usepackage{fullpage}
\usepackage{graphicx}
\numberwithin{equation}{section}
\newcommand{\be}{\begin{equation}}
\newcommand{\ee}{\end{equation}}
\newcommand{\bea}{\begin{eqnarray}}
\newcommand{\eea}{\end{eqnarray}}
\newcommand{\bphi}{\mbox{\boldmath $\phi$}}

\renewcommand{\tilde}{\widetilde}
\renewcommand{\epsilon}{\varepsilon}

\newcommand{\news}{\setcounter{equation}{0}\quad}
\begin{document}
\title{
\begin{flushright}\ \vskip -2cm {\small{\em DCPT-13/39}}\end{flushright}
\vskip 2cm 
A low-dimensional analogue of \\ holographic baryons}
\author{
Stefano Bolognesi and Paul Sutcliffe\\[10pt]
{\em \normalsize Department of Mathematical Sciences,
Durham University, Durham DH1 3LE, U.K.}\\[10pt]
{\normalsize Email: \quad  
s.bolognesi@durham.ac.uk \quad\&\quad\  p.m.sutcliffe@durham.ac.uk}
}
\date{November 2013}
\maketitle
\begin{abstract}
Baryons in holographic QCD correspond to topological
solitons in the bulk. The most prominent example is 
the Sakai-Sugimoto model, where the bulk soliton in 
the five-dimensional spacetime of AdS-type can be approximated by the 
flat space self-dual Yang-Mills instanton with a
small size. Recently, the validity of this approximation has been
verified by comparison with the numerical field theory solution. However, 
multi-solitons and solitons with finite density are 
currently beyond numerical field theory computations. 
Various approximations have been applied to investigate these important
issues and have led to proposals for finite density configurations 
that include dyonic salt and baryonic popcorn.
Here we introduce and investigate a low-dimensional analogue
of the Sakai-Sugimoto model, in which the bulk soliton can be approximated 
by a flat space sigma model instanton. 
The bulk theory is a baby Skyrme model in a three-dimensional 
spacetime with negative curvature. 
The advantage of the lower-dimensional theory is that
numerical simulations of multi-solitons and finite density solutions can
be performed and compared with flat space instanton approximations.
In particular, analogues of dyonic salt and baryonic popcorn configurations
are found and analysed.  
\end{abstract}

\newpage 

\section{Introduction}
\quad\ A common feature of all models of holographic QCD is that baryons 
correspond to topological solitons in the bulk. 
The foremost example of a top-down theory with a string theory
 embedding is the 
Sakai-Sugimoto model \cite{Sakai:2004cn,Sakai:2005yt}, which results
in a Yang-Mills theory with a Chern-Simons term in a five-dimensional
bulk spacetime of AdS-type. As there is an identification between
baryon number and instanton charge, the study of baryons 
is equivalent to the construction of Yang-Mills-Chern-Simons
 solitons in curved space, with a prescribed instanton number. 

At large 't Hooft coupling the Yang-Mills term dominates over the
Chern-Simons term and the soliton has a small size, determined by balancing
curvature and Chern-Simons contributions to the action. As the soliton is
small compared to the curvature scale 
then it can be approximated by a flat space self-dual Yang-Mills 
instanton with a small size \cite{Hong:2007kx,Hata:2007mb}.
The validity of this approximation has recently been confirmed by
numerical field theory computations \cite{Bolognesi:2013nja}, 
by exploiting the $SO(3)$ symmetry of 
the static single soliton to reduce the computation in
four-dimensional space to a reduced theory in a two-dimensional space. 
However, multi-solitons (including solitons at finite density)
are unlikely to have any continuous symmetries, so numerical field theory
computations would require a fully four-dimensional computation that is
beyond current capabilities.

The construction of solitons at finite density is a
crucial aspect for understanding the important issue of dense QCD.
In the limit of a large number of colours, which is the regime of
holographic QCD, cold nuclear matter becomes a crystalline solid, although
the details of this are still to be understood.
It should be possible to capture this behaviour via a bulk soliton 
desription within holographic QCD and in particular within the 
Sakai-Sugimoto model. However, the lack of numerical computations has led to 
various approximate methods being employed to describe this phase, as follows.

Calorons, which are flat space self-dual Yang-Mills instantons 
with a periodic direction, can split into monopole constituents if the period
is smaller than the instanton size. This fact, together with a point
particle approximation, has led to the suggestion \cite{Rho:2009ym}
that the appropriate soliton crystal consists of pairs of dyons with
opposite charges arranged in a salt-like configuration. 
It is argued that, with increasing density, this dyonic salt arrangement 
turns into a cubic crystal of half-instantons that is dual to the well-known
Skyrme crystal.
In a different study, 
making use of approximations involving flat space calorons
 and dilute instantons, it has been proposed 
\cite{Kaplunovsky:2012gb,Kaplunovsky:2013iza}
that with increasing density 
a series of transitions takes place, dubbed baryonic popcorn, where
the three-dimensional soliton crystal develops additional layers in the
holographic direction. Unfortunately, field theory computations are
not yet available to test these ideas in the Sakai-Sugimoto model.  

In this paper we introduce and investigate a low-dimensional analogue
of the Sakai-Sugimoto model. The bulk theory is defined
in a three-dimensional spacetime with negative curvature, and is
an $O(3)$ sigma model with a baby Skyrme term that plays the role of the 
Chern-Simons term in the higher dimensional theory.
It is well-known that instantons in planar sigma models are natural 
low-dimensional analogues of Yang-Mills instantons. If the coefficient
of the baby Skyrme term is small then the soliton has a small size and
may be approximated by an instanton of the flat space sigma model.  
The advantage of the lower-dimensional theory is that
numerical simulations of multi-solitons and finite density solutions can
be performed and compared with predictions using flat space instanton 
approximations. In particular, analogues of dyonic salt and baryonic popcorn 
configurations are found and analysed. The results provide 
evidence to support the validity of these ideas within the 
Sakai-Sugimoto model. 

\section{Solitons of a holographic baby Skyrme model}\news
Consider a $(D+2)$-dimensional spacetime with a metric of the form
\be
ds^2=H
(-dt^2+dx_1^2+\ldots +dx_D^2)
+\frac{1}{H}dz^2,
\ee
where 
\be
H(z)=\bigg(1+\frac{z^2}{L^2}\bigg)^p.
\ee
The warp factor $H(z),$ multiplying the $(D+1)$-dimensional Minkowski
spacetime of the dual boundary theory, 
depends only on the additional holographic coordinate $z.$
The constant $p$ is to be specified later and $L$ determines the curvature
length scale and can be set to unity by an appropriate choice of units.

The metric of the Sakai-Sugimoto model \cite{Sakai:2004cn,Sakai:2005yt}
corresponds to the choice $D=3$ and $p=\frac{2}{3}.$
In this case the spacetime has a conformal boundary as $z\to\infty$ 
and the scalar curvature is
\be
R=-\frac{16(4z^2+3)}{9(1+z^2)^{4/3}},
\ee
with the properties that $R\le 0$ and $R$ is finite (in fact zero)
as $z\rightarrow\infty.$ 

In this paper we are interested in a low-dimensional analogue with $D=1,$
so that the bulk spacetime is three-dimensional with coordinates
$t,x,z.$
In this case, for general $p,$ the scalar curvature is
\be
R=-2p(1+z^2)^{p-2}\bigg((5p-2)z^2+2\bigg).
\ee
For this spacetime to have finite curvature with $R\le 0$ 
requires the restriction $\frac{2}{5}\le p\le 1.$
Later we shall see that a convenient choice is 
$p=\frac{1}{2},$ but for now we consider a general value of $p$
within the above interval.  

The action of the massless $O(3)$ baby Skyrme model in the above spacetime is
\be
S=\int\bigg(
\frac{1}{2}g^{\mu\nu}\partial_\mu\bphi\cdot\partial_\nu\bphi
+\frac{\kappa^2}{4}g^{\mu\nu}g^{\alpha\beta}
(\partial_\mu\bphi\times \partial_\alpha\bphi)\cdot
(\partial_\nu\bphi\times \partial_\beta\bphi)
\bigg)\sqrt{-g}\,dx\,dz\,dt,
\label{action}
\ee
where $\bphi=(\phi_1,\phi_2,\phi_3)$ is a three-component unit vector and
 greek indices run over the bulk spacetime components $t,x,z.$ 
The first term in (\ref{action}) is that of the $O(3)$ sigma model
and the second term, with constant coefficient $\kappa^2,$ is the
baby Skyrme term \cite{Piette:1994ug}.

The associated static energy is 
\be
E=\frac{1}{2}\int \bigg(
\frac{1}{H}\,|\partial_x\bphi|^2
+H\,|\partial_z\bphi|^2
+\kappa^2|\partial_x\bphi \times \partial_z\bphi|^2
\bigg)\,\sqrt{H}\,dx\,dz,
\label{energy}
\ee
and the boundary condition is that $\bphi\to(0,0,1)$ as $x^2+z^2\to\infty.$
As we shall see, this theory has bulk topological solitons 
that share many analogous features to those in the Sakai-Sugimoto model.
The analogue of the baryon number is the integer-valued topological charge
\be
B=-\frac{1}{4\pi}\int \bphi\cdot(\partial_x\bphi \times \partial_z\bphi)
\,dx\,dz,
\label{baryonnumber}
\ee 
which defines the instanton number of the planar sigma model.

Using the fact that the baby Skyrme term 
contribution to the energy is non-negative and $H\ge 1$, together
with the obvious inequality
\be
\bigg|\frac{1}{\sqrt{H}}\partial_x\bphi\pm \sqrt{H}\bphi\times\partial_z\bphi\bigg|^2
\ge 0,
\ee
yields the Bogomolny bound $E\ge 4\pi |B|.$

In flat space ($H=1$) without a baby Skyrme term ($\kappa=0$) this inequality
is attained by the instanton solutions of the $O(3)$ sigma model (for a review
see \cite{Za}).
To write these instanton solutions explicitly it is convenient to use
the equivalent formulation of the $O(3)$ sigma model in terms of the 
$\mathbb{CP}^1$ sigma model. This is obtained by defining the Riemann 
sphere coordinate
$W=(\phi_1+i\phi_2)/(1-\phi_3),$ obtained by stereographic projection of 
$\bphi.$ In terms of this variable, instanton solutions are given by 
$W$ a holomorphic function of $\zeta=x+iz.$ 
The instanton solutions with finite $B>0$ are given by $W(\zeta)$  a
rational function of degree $B,$ where the degree of the numerator is
larger than that of the denominator in order to satisfy the 
above boundary condition. 
Taking into account the global $U(1)$ symmetry associated with the
phase of $W$, this leaves an instanton moduli space ${\cal M}_B$ 
of dimension $4B-1.$
 
The radially symmetric sigma model instanton with topological charge $B$ 
and centre at the origin is given by $W=(\zeta/\mu)^B,$ where 
the positive real constant $\mu$ is the arbitrary size of the instanton. 
Converting back to the $O(3)$ sigma model formulation this 
solution is
\be
\bphi=(\sin f\cos(B\theta),\sin f\sin(B\theta),\cos f),
\label{radialphi}
\ee
where $r,\theta$ are polar coordinates in the $(x,z)$-plane
and $f(r)$ is the radial profile function
\be
f=\cos^{-1}\bigg(\frac{r^{2B}-\mu^{2B}}{r^{2B}+\mu^{2B}}\bigg).
\label{instf}
\ee

If we require that this field configuration has finite energy in the
curved spacetime theory (\ref{energy}) 
then this places a further restriction on the
power $p,$ as follows.
The large $r$ behaviour of the profile function is
\be
f=\frac{2\mu^B}{r^B}+\ldots
\ee
so the greatest restriction arises when $B=1.$
Convergence of the energy requires that $p<2/3,$ which
combined with the earlier restriction yields
$\frac{2}{5}\le p<\frac{2}{3}.$ 
The choice of $p$ is therefore quite
restricted and $p=\frac{1}{2}$ is a natural value to take and will be
used in all our numerical computations. However, we shall often leave
$p$ general in our analytic calculations, to make it clear which terms
derive directly from the curvature of the metric, though it is to 
be understood that any formulae are only to be applied with $p$ in the
above restricted range. 

To make our analogy with the Sakai-Sugimoto model, we consider a regime
in which $0<\kappa\ll 1$; this being the analogue of large 't Hooft coupling.
In this regime the sigma model term dominates over the baby Skyrme term and
it is reasonable to approximate the soliton by a sigma model instanton.
In flat space the static sigma model is conformally invariant and therefore
an instanton has an arbitrary size. However, in the above spacetime
the curvature breaks this energy degeneracy of the sigma model and 
without a baby Skyrme term the 
static energy is a monotonically decreasing function of the instanton size.
The inclusion of the baby Skyrme term produces a contribution to the energy
that increases with decreasing instanton size and balances the curvature
contribution to yield a preferred non-zero instanton size; though it is 
small because the coefficient of the baby Skyrme term is small.
Below we make this qualitative analysis more concrete through explicit
calculation. 
 
To extract the sigma model instanton, that arises in the small $\kappa$ limit,  
we define the rescaled variables $\tilde x=x/\sqrt{\kappa}$ and
$\tilde z=z/\sqrt{\kappa}$, with $\tilde H$ denoting 
$H(\tilde z\sqrt{\kappa}).$ 
In terms of these variables the energy (\ref{energy}) 
becomes
\be
E=\frac{1}{2}\int \bigg(
\tilde H^{-1/2}\,|\partial_{\tilde x}\bphi|^2
+\tilde H^{3/2}\,|\partial_{\tilde z}\bphi|^2
+\kappa \tilde H^{1/2}|\partial_{\tilde x}\bphi \times \partial_{\tilde z}\bphi|^2
\bigg)\,\,d\tilde x\,d\tilde z.
\label{senergy}
\ee
Expanding this energy as a series in $\kappa$ yields
$E=E_0+\kappa E_1+{\cal O}(\kappa^2),$ where
\bea
E_0&=&
\frac{1}{2}\int \bigg(|\partial_{\tilde x}\bphi|^2+|\partial_{\tilde z}\bphi|^2
\bigg)\,\,d\tilde x\,d\tilde z, \label{e0}\\
E_1&=&\int \bigg(
\frac{1}{4}p\tilde z^2(3|\partial_{\tilde z}\bphi|^2-|\partial_{\tilde x}\bphi|^2)
+\frac{1}{2}|\partial_{\tilde x}\bphi \times \partial_{\tilde z}\bphi|^2
\bigg)\,\,d\tilde x\,d\tilde z.
\label{e1}
\eea
The static energy (\ref{e0}) is that of the planar flat space
$O(3)$ sigma model, in 
rescaled variables, and is minimized by the instanton solutions with
$E_0=4\pi B.$ This motivates the approximation of restricting the field
to an instanton configuration and then determining the point in the 
instanton moduli space, ${\cal M}_B,$ that minimizes the energy
(\ref{senergy}). To make analytic progress, a further approximation can
be employed by neglecting terms in the energy beyond linear order in 
$\kappa,$ that is, $E\approx 4\pi B+\kappa E_1.$  With this additional
approximation $E_1$ becomes an energy function on ${\cal M}_B$, and the
task reduces to finding its minimum.

As an illustration of this procedure we first consider a subset of 
${\cal M}_B$ given by restricting to radially symmetric instantons
 $W=(\zeta/\mu)^B,$ with scale $\mu$. The curvature in the holographic
direction acts as an effective potential with a minimum at $z=0$ and
as the energy is invariant under translations in the $x$ direction then
there is no loss of generality in positioning the radial instanton at the
origin. 
In terms of the rescaled variables $\tilde \zeta=\zeta/\sqrt{\kappa}$ and
$\tilde \mu=\mu/\sqrt{\kappa}$ we have $W=(\tilde \zeta/\tilde \mu)^B,$
and for simplicity we first discuss the example with $B=2.$
In this case a simple calculation produces the result
\be
E_1=\pi^2\bigg(p\tilde \mu^2+\frac{4}{\tilde \mu^2}\bigg).
\label{e1tworad}
\ee
The influence of the curvature is evident in the term proportional to $p$ 
in (\ref{e1tworad}) and drives the instanton towards zero size. 
The second term in (\ref{e1tworad}) originates from the baby Skyrme term and 
resists the shrinking of the instanton size.  
 These competing effects combine to produce the finite size 
$\tilde \mu=\sqrt{2}/p^{1/4}$ that minimizes the energy (\ref{e1tworad}).
Returning to the original variables the size is
$\mu=\sqrt{2\kappa}/p^{1/4}$ and is therefore small in the
regime of small $\kappa.$
The energy is dominated by the flat space instanton contribution
$8\pi,$
as the correction from the size stabilizing terms is subleading,
being linear in $\kappa.$

\begin{figure}
\begin{center}
\includegraphics[width=8cm]{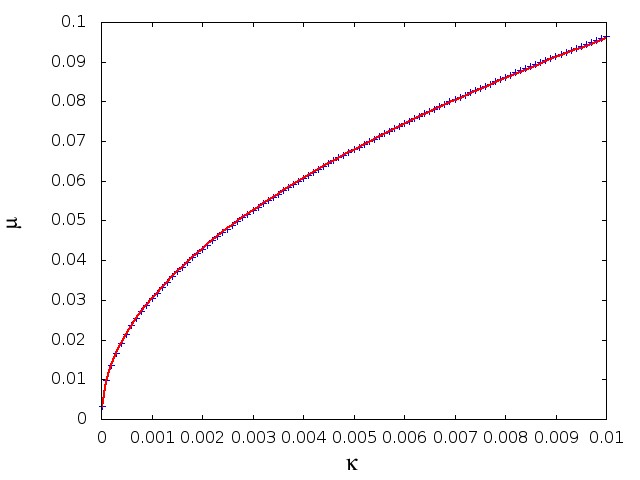}
\caption{The data points give the size $\mu$ of the $B=1$ soliton
as a function of $\kappa,$  within the instanton approximation.
The curve is a numerical fit to the data of the form $c\sqrt{\kappa}$ with
the constant $c=0.96.$
}\label{fig-size1} 
\end{center}
\end{figure}

A similar calculation can be performed for all $B\ge 2$ to show that
the instanton size is proportional to $\sqrt{\kappa}.$
However, the corresponding analytic calculation for $B=1$ is more difficult
because $E_1$ is divergent in this case. The instanton approximation
$W=\zeta/\mu$ is still valid but the series expansion of the energy in terms
of $\kappa$ is not. This is due to the fact that the first order correction 
involves approximating $H(z)$ by a quadratic in $z,$ which is too crude an
approximation for the slow decay of the $B=1$ instanton. 
One approach is to use a more complicated approximation for $H(z)$ that 
captures the correct behaviour to quadratic order but matches better
the large $z$ behaviour of $H(z)$, so that all integrals are finite.
However, this approach is cumbersome and
an alternative method is to numerically compute
the energy (\ref{energy}) of the ansatz $W=\zeta/\mu$ to determine the
minimizing size $\mu$ as a function of $\kappa.$
The result, for $p=\frac{1}{2}$, is displayed in Figure~\ref{fig-size1} for 
$\kappa\in[10^{-4},10^{-2}],$
where the data points correspond to the numerical computation and
the curve is a fit to the data of the form $c\sqrt{\kappa}$ with
the constant $c=0.96.$ This computation suggests that, 
in this regime, the $B=1$ soliton has an approximate size $\sqrt{\kappa},$
in agreement with the above analytic computation for $B=2$ that gives a 
size that is also ${\cal O}(\sqrt{\kappa}).$

The main advantage of our low-dimensional model is that we are able to 
obtain full numerical solutions of the nonlinear field theory, because the
static problem is only two-dimensional and is therefore within the 
reach of modest computational resources. 
All our computations are performed for the model with $p=\frac{1}{2}$
and the parameter value $\kappa=0.01,$ so that the soliton has a size
that is smaller than the curvature length scale. Other smaller values of 
$\kappa$ have also been investigated and
the results are similar.
To numerically compute static solitons we minimize the energy (\ref{energy}) 
using second order accurate finite difference approximations for the spatial
derivatives using lattice spacings $\Delta x=\Delta z=0.005$ and 
numerical grids that contain $2000\times 2000$ points.  
The energy minimization algorithm is an adaptation of the one described in 
detail in \cite{Battye:2001qn}, where it is applied to the three-dimensional
Skyrme model in flat space.

\begin{figure}
\begin{center}
\includegraphics[width=8cm]{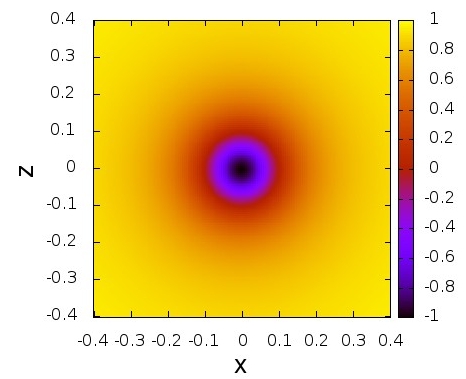}
\includegraphics[width=8cm]{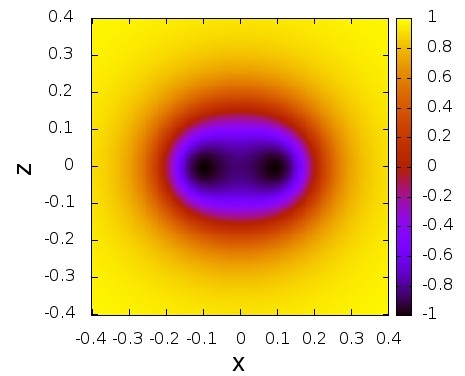}  
\caption{A plot of $\phi_3$ for the soliton with $B=1$ (left image) and
$B=2$ (right image). 
}\label{fig-B12} 
\end{center}
\end{figure}

The result of our field theory computation for the $B=1$ soliton is
displayed in the left image in Figure~\ref{fig-B12}, where we plot
$\phi_3,$ as this gives a good pictorial representation of the soliton.
It can be seen that the soliton has an approximate radial symmetry and a
size that is roughly $\sqrt{\kappa}=0.1,$ as predicted by the 
instanton approximation. In fact, a plot of $\phi_3$ using the
instanton approximation $W=\zeta/\mu$ with a size $\mu=0.96\sqrt{\kappa}$
produces an indistinguishable image. The accuracy of the instanton
approximation is confirmed by an energy comparison, as the 
numerical field theory computation yields a value 
$E=4\pi \times 1.0148$ to be compared with the 
energy of the instanton approximation $E=4\pi \times 1.0154$.

The right image in Figure~\ref{fig-B12} displays the result of the
field theory computation for the $B=2$ soliton. The energy is
$E=8\pi \times 1.0105,$ so this is a bound state of two single solitons.
Note that the binding energy per soliton is less than $0.5\%$ of the
single soliton energy, illustrating the fact that holographic models
based on perturbations around BPS systems can yield the kind of 
small binding energies found in real nuclei. 
The $B=2$ soliton does not have (even approximate) radial symmetry but
resembles two single solitons separated along the non-holographic direction,
with a separation that is close to the diameter of a single soliton. 
Below we present an analytic calculation that explains this result
by making use of the 2-instanton moduli space ${\cal M}_2.$

\begin{figure}
\begin{center}
\includegraphics[width=10cm]{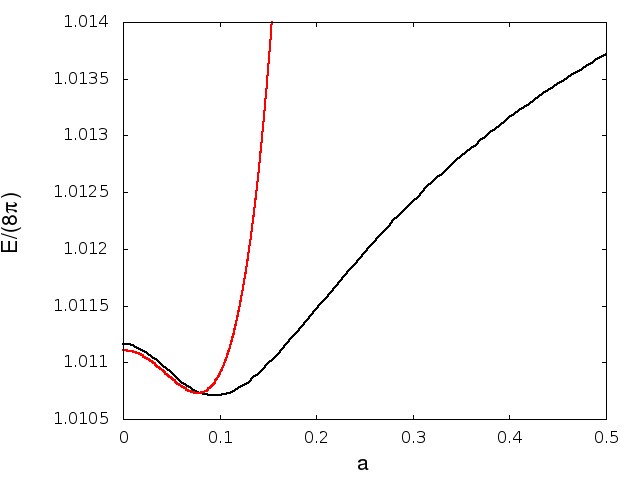}
\caption{A plot of $E/(8\pi)$ for a 2-instanton configuration
with separation $2a$ between the two instantons.
The black curve is the result of a numerical integration to compute the
 energy and the red curve is the analytic approximation described in the
text. 
}\label{fig-pot} 
\end{center}
\end{figure}

We investigate the $B=2$ soliton by considering a two-dimensional subset of
${\cal M}_2$ that includes radial instantons. Explicitly, we consider 
instanton solutions of the form 
\be
W=\frac{\zeta^2-a^2}{\mu^2},
\label{charge2}
\ee
where $\mu$ and $a$ are real parameters. This describes two instantons
separated along the non-holographic direction 
with positions $(x,z)=(\pm a,0)$ and $\mu$ determines their equal size.
Radial solutions correspond to the choice $a=0.$
The idea is to determine, for each $a$, the size $\mu$ that minimizes
the energy (\ref{energy}) and then show that as a function of $a$ this 
energy has a minimum at a value of $a$ that is  ${\cal O}(\sqrt{\kappa}).$
For any given value of $\kappa$ this calculation can be performed by
numerical integration of the energy (\ref{energy}). As an example, for
$\kappa=0.01$ and $p=\frac{1}{2}$ the black curve in Figure~\ref{fig-pot}
displays $E/(8\pi)$ as a function of $a,$ which clearly has a minimum
at $a\approx 0.1=\sqrt{\kappa}.$ 

To make analytic progress we perform the same rescaling introduced earlier,
writing $\tilde \mu=\mu/\sqrt{\kappa}$ and $\tilde a=a/\sqrt{\kappa}$,
and evaluate the first order correction to the energy $E_1.$
The resulting integrals can be evaluated by expanding around the radial
case $\tilde a=0$ to extend the result (\ref{e1tworad}).
To quartic order in $\tilde a$ 
 \be
E_1=p\bigg(\pi^2\tilde\mu^2-2\pi\tilde a^2+\frac{\pi^2\tilde a^4}{\tilde\mu^2}\bigg)
+\frac{4\pi^2}{\tilde\mu^2}+\frac{5\pi^2\tilde a^4}{\tilde \mu^6},
\label{e1scaled}
\ee
so the minimizing size is
\be
\tilde\mu=\frac{\sqrt{2}}{p^{1/4}}+\frac{1}{2^{3/2}}p^{3/4}\tilde a^4.
\ee
Substituting this size into (\ref{e1scaled}) produces 
\be
E_1=4\pi^2\sqrt{p}-2\pi p\tilde a^2+\frac{3}{4}\pi^2p^{3/2}\tilde a^4,
\ee
hence the radial instanton is unstable to a perturbation that separates
the two instantons along the non-holographic direction. Approximating
the energy by neglecting quadratic and higher order terms in $\kappa,$
that is $E\approx E_0+\kappa E_1,$ and returning to unscaled variables,
the above result yields
\be
\frac{E}{8\pi}\approx 1+\frac{\pi}{2}\sqrt{p}\kappa-\frac{1}{4}pa^2
+\frac{3\pi}{32\kappa} p^{3/2} a^4.
\label{eapprox}
\ee
This energy is minimized at the value
\be
a=\frac{2}{p^{1/4}}\sqrt\frac{\kappa}{3\pi},
\ee
so the required ${\cal O}(\sqrt{\kappa})$ dependence has indeed been 
obtained. 
Furthermore, 
this calculation reveals that the separation $2a$ between the two instantons 
is approximately equal to the size $\mu$ of the 2-instanton. In other words,
it predicts that the 2-soliton solution should closely resemble
 two touching single solitons.

To test the accuracy of this approximation we can compare
the formula (\ref{eapprox}) to the earlier numerical computation 
(black curve in Figure~\ref{fig-pot})
with $\kappa=0.01$ and $p=\frac{1}{2}.$ 
For these parameter values the approximation
(\ref{eapprox}) is shown as the red curve in 
Figure~\ref{fig-pot}.
This shows that the qualitative features are correct and that the 
approximation is reasonably accurate for separations up to the 
minimizing value, and in particular the estimate for the energy
minimizing separation is reasonable. 

From the numerical data used to produce the
black curve in Figure~\ref{fig-pot}, it is found that the 
energy is minimized for the instanton parameters
$a=0.095$ and $\mu=0.18$ with an energy that is only 
$0.03\%$ above the field theory computation.
As expected from this result, a plot of $\phi_3$ using this
instanton approximation produces an image that is 
essentially identical to the right image displayed in Figure~\ref{fig-B12}.

We have also computed solitons with larger values of $B$.
The result yields a string of $B$ single solitons that are touching and aligned 
along the non-holographic direction, as expected given
the above result for $B=2$ and the fact that the curvature favours solitons
located along the line $z=0.$ 
In the following section we turn our attention to solitons with finite
density and display some interesting phenomena.

\section{Solitons at finite density}

\quad\ As mentioned in the introduction, the study of solitons at 
finite density in the Sakai-Sugimoto model has attracted some recent attention
in attempts to understand dense QCD within a holographic setting.
However, various levels of approximation have to be made to make any progress
on this topic, given that numerical field theory simulations are currently not
available. Furthermore, even the relevant flat space self-dual instanton
solutions are not explicitly known once periodic boundary conditions
are applied in multiple directions, which is precisely the case of interest.  
However, in our low-dimensional analogue not only
are we able to compute the
numerical field theory solutions, but we also have access to simple explicit
formulae for the relevant periodic flat space instantons. This allows us
to investigate solitons with finite density using both approaches and to 
compare the results. In particular, we find analogues of the phenomena 
suggested for the Sakai-Sugimoto model and are able to study these in some
detail.

To numerically compute solitons at finite density we restrict our numerical
grid in the non-holographic direction to the range $-L\le x\le L$ and impose
periodic boundary conditions by identification of the fields at $x=\pm L.$
The integral expression (\ref{baryonnumber}) for the baryon number, 
with the range of integration now
restricted to the strip $(x,z)\in[-L,L]\times (-\infty,\infty),$ is still
integer-valued and defines the finite density $\rho=B/(2L).$  
Initial conditions are obtained by using the radial field (\ref{radialphi})
with $B=1$ and a profile function $f(r)$ with compact support so that
$f(0)=\pi$ and $f(r)=0$ for $r\ge r_{\star},$ 
for a suitably chosen constant $r_{\star}.$
A periodic field with 
topological charge $B$ can then be created by a linear superposition of 
$B$ fields of this form, where the position of each soliton is shifted
so that the separation between any pair of solitons is greater than
$2r_{\star}.$ In this construction of the initial field configuration, 
each soliton can independently be given an arbitrary phase 
$\chi,$ corresponding to
a shift $\theta\mapsto\theta+\chi$ in the radial field (\ref{radialphi}).

In fact, we already have evidence that the most attractive arrangement of
two solitons is associated with a relative phase of $\pi$ between the
two solitons. This evidence follows from the following argument.
In terms of the $\mathbb{CP}^1$ coordinate $W,$ the product ansatz for 
two fields $W_1$ and $W_2$ is 
\be
W=\frac{W_1W_2}{W_1+W_2}.
\label{prod}
\ee
This has the property that $W$ is zero whenever $W_1$ or $W_2$ is zero,
so there are solitons at the positions given by the two 
constituent components. 
Furthermore, in a region of space far from the soliton 
described by one of the components, say $W_2,$ then this component is
large so $W\approx W_1,$ and vice-versa. Consider the above product
for two solitons of equal size, with a relative phase $\chi$ and 
positions $\pm a$ along the $x$-axis. Explicitly, 
\be
W_1=\frac{\zeta-a}{\nu},\quad
W_2=e^{i\chi}\frac{\zeta+a}{\nu},\quad
\mbox{producing}
\quad
W=\frac{e^{i\chi}(\zeta^2-a^2)}{\nu(\zeta(e^{i\chi}+1)+a(e^{i\chi}-1))}.
\ee
For the above field to match the expression (\ref{charge2}), that is
a good description of the mininal energy $B=2$ soliton, requires
$e^{i\chi}=-1$ and $\nu=\mu^2/(2a).$ Thus to 
have the two solitons in the most attractive channel requires
the relative phase to be $\pi,$ that is neighbouring solitons should be
exactly out of phase.

The above argument is confirmed by our numerical simulations at finite
density, where it is found that a chain of solitons which are all in 
phase has a higher energy per soliton than a chain in which the 
phase of the solitons
alternates between 0 and $\pi.$ Note that in the case of alternating phases
the number of solitons in the chain in a fundamental period must be even to 
have a strictly periodic field. 
Under translation by the inter-soliton separation  
the field is only periodic up to a change of sign.

\begin{figure}
\begin{center}
\includegraphics[width=8cm]{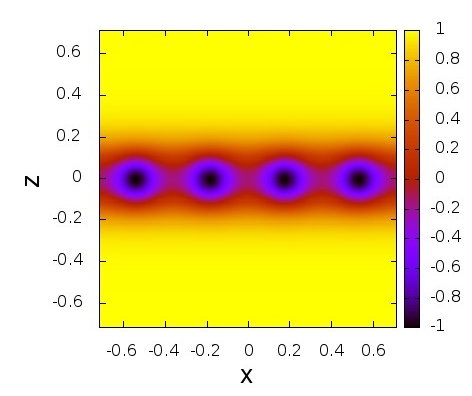}
\includegraphics[width=8.2cm]{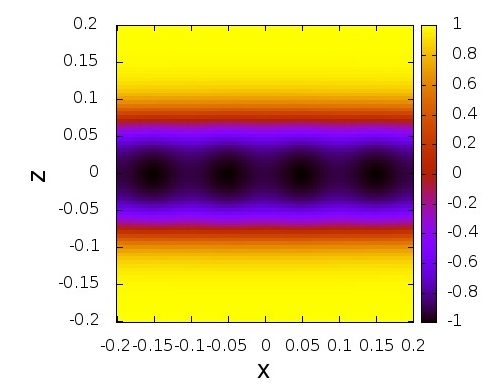}  
\caption{A plot of $\phi_3$ for a chain with the optimal density 
$\rho=2.8$ (left image) and the density $\rho=10$ (right image).
}\label{fig-best} 
\end{center}
\end{figure}

\begin{figure}
\begin{center}
\includegraphics[width=11cm]{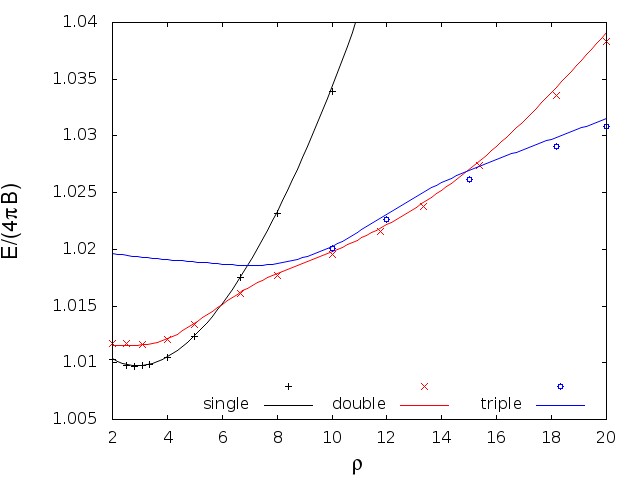}
\caption{
A plot of $E/(4\pi B)$ against density $\rho$ for the single chain, 
double chain and triple chain configurations. 
Data points are the numerical solutions from field theory simulations
and curves are sigma model instanton approximations.
}\label{fig-chain} 
\end{center}
\end{figure}

To calculate the energy per soliton $E/B$ as a function of the density 
$\rho=B/(2L)$ for a single chain of
solitons with alternating phases, we perform computations for a range of
grid sizes $2L,$ using an initial configuration with $B=4,$ so that
the simulation region contains two fundamental periods.
The optimal density is found to be $\rho=2.8,$ at which
the energy per soliton is
 $E/B=4\pi\times 1.0097.$  The left image in Figure~\ref{fig-best} 
displays $\phi_3$ for this optimal density chain.
In Figure~\ref{fig-chain} the data points marked with a $+$ represent
$E/B,$ in units of $4\pi,$ for the numerical field theory solution of
this chain. Note that the optimal density corresponds to the critical
value of the chemical potential at which there is a 
first order phase transition to an equilibrium density of solitons,
this being the analogue of the nuclear matter phase transition in QCD. 

The chain solution at high density is displayed
in the right image in Figure~\ref{fig-best}, which corresponds to the
density $\rho=10,$ being more than three times the optimal density. 
At such a high density each soliton splits into a kink anti-kink pair
separated along the holographic direction, and the solitons lose their
individual identities to form an almost homogeneous structure in the
non-holographic direction. This is the lower-dimensional analogue of the
appearance of monopole constituents for calorons and has been discussed
previously for instantons of the $O(3)$ sigma model in flat space 
\cite{Bruckmann:2007zh,Eto:2006mz,Harland:2009mf}.
A configuration of this type is therefore our low-dimensional analogue
of the dyonic salt arrangement discussed in \cite{Rho:2009ym}.
Furthermore, we can make use of the flat space sigma model results to
study finite density configurations semi-analytically, as follows.

The starting point is the periodic sigma model solution 
considered in \cite{Bruckmann:2007zh}
\be
W=\nu\sin(\pi\rho\zeta),
\label{schain1}
\ee
that describes an instanton chain in which there are instantons located along
the $x$-axis with a distance $1/\rho$ between neighbouring instantons 
(recall that the position of an instanton corresponds to a point
where $W$ vanishes). Note that this field obeys the symmetry relation
$W(\zeta+\rho^{-1})=-W(\zeta),$ which shows that neighbouring instantons are
exactly out of phase, so this is a chain of instantons with alternating phases.
The real parameter $\nu$ controls the size of each instanton, which is given
by $1/(\nu\pi\rho)$ in the dilute regime where the instanton size is 
small compared to $1/\rho.$ Once the instanton size is comparable 
to $1/\rho$ the instantons lose their individual identities and as
the size increases further they split into kink anti-kink constituents
and the configuration tends towards a homogeneous state
\cite{Bruckmann:2007zh,Eto:2006mz,Harland:2009mf}.

The above instanton chain can be used to approximate the soliton
chain of our holographic model.
For any given density $\rho,$ the energy 
(\ref{energy}) of the field (\ref{schain1}) can be computed
by numerical integration and a minimization performed over the parameter $\nu$ 
to determine the minimal energy. The result of such a 
computation is displayed as the black curve in Figure~\ref{fig-chain},
where it can be seen that this instanton approximation is in excellent 
agreement with the data from field theory simulations.
We have therefore shown, using two different approaches, how a 
chain of solitons splits into kink anti-kink constituents, in analogy
to the proposed dyonic salt. 

\begin{figure}
\begin{center}
\includegraphics[width=8cm]{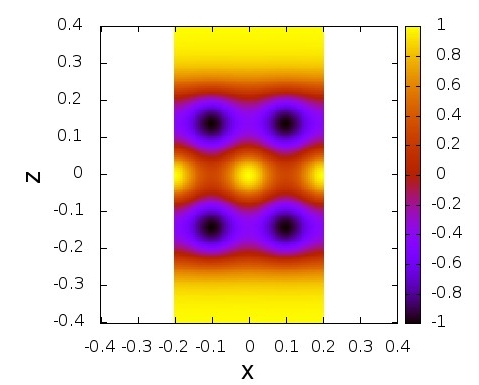}
\includegraphics[width=7.8cm]{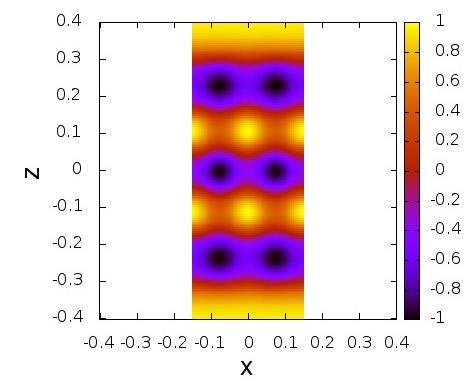}
\caption{A plot of $\phi_3$ for the double chain 
solution with density $\rho=10$ (left image) and the
triple chain solution with density $\rho=20$ (right image).
}\label{fig-dt} 
\end{center}
\end{figure}

Beyond the optimal density, the energy per soliton of the chain grows 
rapidy with increasing density. The baryonic popcorn idea introduced in
\cite{Kaplunovsky:2012gb,Kaplunovsky:2013iza} suggests that there will be
a critical density beyond which it is energetically preferable for the soliton
chain to split into a pair of chains via a pop into the holographic
direction. To investigate this possibility we perform field theory 
simulations using initial conditions that describe a double chain.
The result for the double chain with density $\rho=10$ is displayed in
the left image in Figure~\ref{fig-dt}, where the phase of the
solitons alternates within each chain.
 The two chains are aligned and two solitons in 
different chains with the same $x$ coordinate are out of phase.
This is the expected arrangement to allow maximal attraction between
all neighbouring solitons and it can be confirmed that other possibilities,
such as a relative shift between the two chains or a zig-zag arrangement,
relax back to this solution under energy minimization.
In Figure~\ref{fig-chain} the data points marked with a $\times$
denote the energy per soliton (in units of $4\pi$) as a function of 
the density 
for the double chain solution obtained from field theory simulations.
These results show that the double chain solution has a lower energy
than the single chain solution once the density is greater than 
about twice the optimal density.

As in the case of the single soliton chain, the double soliton chain
can also be studied using an instanton approximation. The appropriate 
instanton can be obtained using the product ansatz (\ref{prod}) with
constituent fields obtained by translation of the single chain form
(\ref{schain1}). Explicitly, take
\be
W_1=\nu\sin({\pi}\rho(\zeta-i\delta)/2), \quad\quad
W_2=-\nu\sin({\pi}\rho(\zeta+i\delta)/2),
\label{schain2} 
\ee
and minimize the energy over the scale $\nu$ and the distance $2\delta$
between the chains in the holographic direction. The result is 
shown as the red curve in Figure~\ref{fig-chain} and is again in good
agreement with the field theory computations.
Both approaches therefore confirm the analogue of baryonic popcorn
in our low-dimensional model. In particular, in the low-dimensional theory
we find that the popcorn phenomenon takes place at a density below that at 
which the solitons split into constituents, so our popcorn is not salted. 

In the study of baryonic popcorn in the Sakai-Sugimoto model it has been
proposed that multiple chains have a zig-zag structure
\cite{Kaplunovsky:2012gb,Kaplunovsky:2013iza}. 
A zig-zag structure requires that the optimal separation for two solitons
is much greater than the size of a single soliton. As this is not the case
in our low-dimensional theory then it is not surprising that zig-zag 
solutions do not appear. The double chain instanton (\ref{schain2})
can be modified to describe a zig-zag structure and this also
supports the conclusion that it
is not an energetically favourable configuration in this theory.

As one might expect, as the density is increased further the lowest energy
configuration becomes a triple chain solution. An example field theory
computation for the density $\rho=20$ is shown in the right image in 
Figure~\ref{fig-dt}. The energy per soliton (in units of $4\pi$) 
for triple chain solutions computed from field theory simulations are
shown in Figure~\ref{fig-chain} as the data points marked with a 
$\circ$ for several densities. A triple chain instanton can be 
obtained by an obvious generalization of the double chain instanton
(\ref{schain2}). A minimization over the parameters in such an
instanton approximation yields the blue curve in Figure~\ref{fig-chain},
which agrees with the field theory results.
These results confirm that a triple chain solution is energetically
preferred over a double chain solution for sufficiently high density,
as expected from the popcorn phenomenon. All these results suggest that
as the density is increased further then the number of soliton chains
increases and eventually the configuration begins to resemble a portion of
a two-dimensional lattice rather than the one-dimensional chain that 
arises at the optimal density. 
This is exactly the phenomenon predicted in \cite{Kaplunovsky:2012gb,Kaplunovsky:2013iza}.

\section{Conclusion}
\quad\ In this paper we have introduced and investigated a 
holographic baby Skyrme model that can be used to study several
low-dimensional analogues of bulk solitons in the Sakai-Sugimoto model,
describing holographic baryons. 
The advantage of the low-dimensional theory is that several aspects 
that one would like to study in the Sakai-Sugimoto model, but are currently 
not tractable, can be investigated exactly. 
In particular, we have compared sigma
model instanton approximations with the results of field theory simulations
and shown that this provides a good approximation for solitons, 
multi-solitons and finite density solutions. 
This provides further support for the use of self-dual Yang-Mills instantons
in approximating bulk solitons in the Sakai-Sugimoto model.  
Analogues of dyonic salt
and baryonic popcorn configurations have been found, providing strong
evidence for their relevance in the study of holographic baryons. 

We have used a baby Skyrme term to provide a low-dimensional analogue
of the Chern-Simons term that appears in the Sakai-Sugimoto model.
Perhaps a more obvious analogue of the Chern-Simons term is to couple
a vector meson to the sigma model 
topological current, as studied in flat space in \cite{Foster:2009rw}.
Note that the results in that paper show that this can produce similar 
results to the inclusion of a baby Skyrme term. As the baby Skyrme model
is easier to investigate we have chosen this route in the current paper, but
it might be of interest to investigate the vector meson model to see if
additional phenomena can be obtained. 

\section*{Acknowledgements}
This work is funded by the EPSRC grant EP/K003453/1 and 
the STFC grant ST/J000426/1.

\end{document}